\documentclass[10pt,conference,a4paper]{IEEEtran}
\usepackage{times,amsmath,epsfig}
\title{Electrically small metamaterial-based antennas -- have we seen any real practical benefits?}
\author{%
{Pekka Ikonen}
\vspace{1.6mm}\\
\fontsize{10}{10}\selectfont\itshape
Nokia Devices R\&D Productization Technology Management\\
P.O.Box 407, 00045 NOKIA GROUP, Finland\\
\fontsize{9}{9}\selectfont\ttfamily\upshape
pekka.1.ikonen@nokia.com
}
\begin{document}
\maketitle
\begin{abstract}
Electrically small metamaterial-based antennas are discussed from the industrial point of view using mobile phones as the application example. It appears, that despite the interesting theoretical findings, the commercial acceptability of these antennas is low. Some of the issues possibly leading to this situation are addressed. Discussion topics range from challenging application environment, through the response of finite-size composite-material samples, all the way to the required constructive criticism and acknowledgement of prior art. Selected issues are discussed in more details, and proposals how to possibly improve the commercial acceptability of metamaterial-based antennas are made.
\end{abstract}


\section{Introduction}

The number of papers about electrically small metamaterial-based antennas is big and steadily growing, e.g.~\cite{Ziolkowski03, Mahmoud04, Tretyakov05, Qureshi05, Itoh06, Stuart06, Ziolkowski06, Alu07, Ziol07, Mosallaei08, Erentok08, Hirvonen08, Bilotti08}, and the references therein. Interesting theoretical discussions predict great advantages from metamaterials in small-antenna design. For example, resonant conditions for strongly subwavelength patch antennas, or possibilities to overcome the small-antenna $Q-$limit have been discussed.

Undoubtedly, metamaterial-inspired theoretical ideas can offer new points of view in the ``traditional'' small-antenna design. Also, many of the theoretical works are already backed up with prototypes experimentally verifying the proposed ideas. However, to push the proposed antennas in commercial applications (e.g., in mobile phones), it is necessary to properly demonstrate the practical benefits of metamaterial-based antennas when compared with ``traditional'' reference antennas for the same application. Unfortunately, at the time being, solid comparative demonstrations can hardly be found in the literature.

What is the root cause for the lack of these demonstrations? Understanding this is very important as convincing experimental demonstrators are essential to maintain the industrial interest in small-antenna enhancement using metamaterials. Below we discuss some challenges for the utilization of metamaterials in mobile-phone antennas, and address some other issues that might be hindering the commercial acceptability of these design schemes. Previously, some challenges related to metamaterials in small-antenna design have been discussed, e.g.,~in \cite{Mittra1, Ikonen_pres}.

\section{Metamaterials in mobile-phone antennas:~some general observations}

To understand better how we could possibly push more metamaterial-based (-inspired) antennas to mobile phones, we start by listing down some related general challenges. Later, some of the items listed below are discussed in more details, and proposals for clarification practices are made. At the end, the goal is to ensure that always the most competitive antenna finds its way to the target application.\\ \\
\underline{1.} Mobile phone is a very challenging application environment for artificial composite materials.
\begin{itemize}
\item Available volume in the vicinity of antennas is only a small fraction of free-space wavelength.
\item Antenna manufacturing complexity should be kept low.
\item Spatial near fields exciting finite-size material samples are highly complex.
\item Some of the antennas need to couple strongly enough to the rest of phone mechanics to gain sufficient bandwidth.
\end{itemize}
\underline{2.} Resonant nature of typical metamaterials is challenging.
\begin{itemize}
\item Interesting material phenomena tend to occur in the vicinity of the resonance.
\item Effect of dispersion and resonant losses can be strong.
\item Non-radiating resonances coupled with antenna resonances are unwanted.
\end{itemize}
\underline{3.} Metamaterial-based antennas are rarely fully benchmarked against reference antennas.
\begin{itemize}
\item Full experimental characterization in the real application environment is always needed.
\item Reference antenna should be targeted (desirably already being used) for the same application.
\item Proper figure-of-merit should be used:~bandwidth comparison is not enough if efficiency degrades.
\end{itemize}
\underline{4.} Occasionally self-driven constructive criticism towards metamaterial-based antennas is missing.
\begin{itemize}
\item Why should the proposed antenna be actually used, instead of ``traditional'' antennas?
\item Possible drawbacks (increased weight, cost etc.) of the proposed designs should be openly stated.
\item The realized response of metamaterials is always restricted by fundamental physical laws $\rightarrow$ these restrictions should always be stated even with theoretical works.
\item Also ``not-optimally-working'' antenna designs might be valuable, if the study increases physical understanding.
\end{itemize}
\underline{5.} Proper acknowledging of prior works and proper marketing is needed.
\begin{itemize}
\item Artificial materials in microwave engineering have a long history.
\item Practical realizations of metamaterial-antennas often resemble ``traditional'' antennas. When should one market the solution as ``metamaterial-based antenna''?
\item Metamaterial-based antennas should be marketed as alternative solutions to traditional antennas, rather than the only possible solution.
\item Consistent terminology derived from prior works should be used $\rightarrow$ ``traditional'' antenna features should not be hidden behind new terminology.
\end{itemize}

\section{Mobile-phone antennas as the application for metamaterials}

The largest dimensions of a mobile phone are roughly $\lambda_0/3...\lambda_0$ over the commonly used communications frequencies ($\lambda_0$ is the free-space wavelength). The volume reserved, e.g., for the cellular antenna is therefore only a small fraction of the wavelength. In addition to this, the spatial near-field profile in the vicinity of the antenna is typically highly complex due to the antenna pattern details, and closely located mechanics components (display, speakers, etc.) Thus, it is impossible to create the ideal homogenization conditions assumed in many theoretical works. Also, due to the very small volume reserved for the antenna, the whole phone is typically utilized as a radiator in order to increase the obtainable bandwidth \cite{Vainikainen}. Apparently, the difference between the response (even the goal of the desired response) of a free-standing antenna element, and the element mounted in a real mobile phone can be significant. To promote the findings successfully from the commercial point of view, it is therefore essential to make sure that the proposed antenna offers the best size-vs.-performance characteristics also in the real phone environment.

To maintain low manufacturing complexity (and associated costs), a big portion of mobile phone antennas is still implemented on planar surfaces. Even though 3D composite material covers (e.g.~\cite{Ziolkowski03, Stuart06, Mosallaei08}) would allow (in theory) to obtain natural matching for a highly sub-wavelength antenna, it is difficult to envision the actual realization of such covers in low-volume and low-cost applications. At the end, the performance enhancement obtained even with planar substrates under volumetric antenna elements (e.g., planar inverted F-antenna) should clearly outweigh the increased manufacturing process complexity (costs), increased weight, and implications of the reserved volume.

\section{On the resonant nature of typical metamaterials}

Typical realizations of metamaterials proposed for electrically-small antennas are composite substrates or superstrates based on resonant inclusions. For example, Lorentz-type resonant magnetic behavior is achieved with a lattice of broken loops, and Drude-type artificial permittivity behavior is achieved with a lattice of thin wires. Alternatively, transmission-line meshes can be used to create high$-k$ ($k$ is the propagation constant in the mesh) appearance for a wave oscillating in the mesh with the goal to obtain size reduction.

Even when excluding the above described challenges related to the mobile-phone volume constraints, there remain some fundamental questions on other challenges. For example, artificial magnetics have no natural magnetic polarization, thus, work has to be done to polarize the loops to obtain collective microwave response. Moreover, the loops are electrically very small, thus, their contribution to total radiation is typically negligible. Rather, the loops tend to store energy in the near field around them. How could this kind of material help boosting the performance of the main radiator whose main loss mechanism should come through radiation?

In general, typically the exotic, ``metamaterial-like'' phenomena occur in the vicinity of the material resonance, thus, such a material possesses strong dispersion and resonant losses. Coupling this kind of materials with inherently rather high$-Q$ antennas creates some apparent challenges: strong dispersion further increases the antenna $Q$ (most often undesirable, example discussion in \cite{Ikonen}), or a discrete collection of inclusions acts more as a non-radiating parasitic resonator than a ``true'' material load \cite{IkonenMOTL}. In the latter case, it sometimes becomes difficult to identify the differentiating advantage offered by metamaterial-based antenna implementations when compared to ``traditional'' solutions utilizing parasitic resonators to boost the bandwidth. Moreover, due to very tight system requirements for the radio performance, non-radiating resonances only boosting the impedance bandwidth (and not the radiation efficiency bandwidth) are typically unwanted.

\section{Proper experimental antenna performance benchmarking is essential}

How to get new antenna concept adopted in commercial use, e.g., in mobile phones? First, the benefits (smaller volume or improved performance with a fixed volume, etc.) stemming from the proposed solution should clearly enough outweigh the possibly associated challenges (increased weight, complexity and cost, etc.). Second, given the performance of the proposed antenna seems feasible, this performance should be compared with the performance of a reference antenna being used for the particular application. Below we list some general issues that help to build a convincing demonstration.\\ \\
1. The reference antenna is properly chosen.
\begin{itemize}
\item The reference antenna should preferably already be used for the proposed application.
\item Several different antennas are being used in mobile phones $\rightarrow$ it is most convincing to compare the proposed antenna to the most competitive available antenna.
\end{itemize}
2. The antennas are experimentally characterized in proper environment.
\begin{itemize}
\item Antennas targeted to mobile phones should be characterized over proper-size chassis.
\item Value of the results is increased if real mechanics components close to the antenna (battery, speakers, etc.) are included to the printed-wiring-board prototype.
\item Most convincing demonstrations are obtained with real phone mechanics (using existing phones).
\end{itemize}
3. The antennas are completely characterized.
\begin{itemize}
\item Only the absolute value of $S_{11}-$parameter is clearly a non-complete description of small-antenna performance.
\item Measured efficiency and input impedance behavior should be presented.
\item Value of the results is further increased by considering also the user effect on the antenna performance.
\end{itemize}
4. Proper figure-of-merit is used in the performance comparison.
\begin{itemize}
\item For single-resonant antennas proper figure-of-merit describing size-vs.-radiation bandwidth characteristics is the radiation quality factor.
\item For multi-band antennas, possibly accompanied with a matching circuit, determining a proper figure-of-merit becomes more challenging.
\item Often in these cases performance has to be evaluated as a compromise between required volume, impedance behavior, total efficiency, tolerance effects of matching components, tolerance to user effects, and manufacturing complexity and cost.
\end{itemize}
5. Both the benefits and drawbacks of the proposed solution are fully reported.

\section{Self-driven constructive criticism towards the proposed solutions}

The world is full of differently seeming electrically small antennas. Evidently, a lot of attention has been paid to the selection of certain antennas for the use in mobile phones. Some of the issues typically affecting this selection process have been described above. Therefore, as metamaterial-based antennas are being proposed for mobile phones, the proposal should first clearly answer to the question: ``Why should the proposed antenna be used over all the other alternatives?''

Especially in the beginning of metamaterial research these materials were in many occasions advertised to provide characteristics not found in nature. Such advertisements, accompanied with some first theoretical studies based on simplified material models, have created a lot of expectations towards metamaterials also in the field of small antennas. It is apparent, however, that as we approach the experimental realization of antennas utilizing these materials, inevitable performance restrictions (e.g., dispersion and losses) are strongly limiting the actual performance. Therefore it is important to understand and openly state the practical limitations even in the case of (typically the first) most theoretical studies, not to create hypothetical expectations.

For example, for some time artificial magnetic materials were considered as a very good miniaturization technique for microstrip antennas due to the low-loss nature of the corresponding microwave (artificial) magnetism (background for magnetic materials with microstrip antennas is available, e.g., in \cite{Hansen}). However, the experimental demonstrations available in the literature were incomplete, or failed to validate the observations based on simplified analysis (see \cite{Ikonen} and the reference therein for more discussion). When the inherent material dispersion (coming as an inevitable side result of the experimental realization) was included into the analysis, it was revealed that such materials can never outperform reference antennas \cite{Ikonen}.

\section{Acknowledging prior works and proper marketing}

The history of artificial materials in microwave engineering is very long, especially when it comes to to the utilization of artificial dielectrics (see, e.g.,~\cite{Brown} for a collection of related early references). Also, the transmission-line and resonator theories have been well established for several decades ago. Thus, as already outlined above, some of the realizations of metamaterial-based antennas might bear strong resemblance with the ``traditional'' solutions. However, still in this case the proposed antennas might offer some benefits not seen in the prior solutions. Nevertheless, when introducing the proposed antennas it is important to understand and respect the prior works, to be able to clearly highlight the differentiating aspects of the proposed solution.

An illustrative example of a good practice is the case of negative permittivity resonator (sphere antenna) \cite{Stuart06}. This antenna seems to bear a striking resemblance with the spherical helix resonator introduced by Wheeler 50 years ago \cite{Stuart06, Wheeler}. Nevertheless, the example shows how a metamaterial-inspired theoretical idea materialized into an interesting antenna design, and one of the contributors explicitly acknowledged the resemblance to the prior works \cite{Stuart_conf}.

Other issues possibly helping to improve the commercial acceptability of metamaterial-based antennas through better understanding relate to using consistent terminology. Currently, confusion is created as occasionally non-standard evaluation measures are used (for related criticism see, e.g.,~\cite{Kildal06}), or widely studied structures are called differently in different sources. For example, despite the different terminology being used, all the structures considered in \cite{Bilotti08, Ikonen, Mosallaei07} physically boil down to a periodic array of broken loops (authors of \cite{Buell06} further call a principally similar substrate ``magnetic metamaterial'' substrate). A reader not experienced with the progress in this field might have the illusion that different structures are studied in all of these papers.

It is also common that many antenna structures available in the recent literature are called ``metamaterial-based antennas'' or simply ``metamaterial antennas'', even though the actual structures do not contain anything that can be described as (artificial) material according to general definition \cite{Sergei_book}. Examples of such antennas include, e.g., microstrip antennas utilizing only one discrete resonant grid as a superstrate, or antennas utilizing few discrete resonators (often broken loops) within the antenna volume. For many people having background in the field of small antennas (but not necessarily in the field of metamaterials) the use of such terminology might create the feeling of an attempt to hide traditional antenna features behind newly established terminology.

\section{Some concluding remarks}

Electrically small metamaterial-based antennas have been briefly discussed from the industrial point of view using mobile phones as the application example. We have listed down several issues possibly affecting the fact that, despite interesting theoretical findings, the commercial use of metamaterial-based antennas, e.g., in mobile phones is low. Some of the issues, like the challenging application environment, cannot be affected. Other issues, like the proper experimental characterization of the proposed antennas, will have a clear impact when trying to push these antennas to commercial applications. Also, it has been highlighted that one has to be constructively critical towards the proposed antennas, as this, added to proper experimental characterization and acknowledgement of prior art, is the best way to ensure that at the end the most competitive antenna finds its way to the target application.

\bibliographystyle{IEEEtran}

\bibliography{IEEEabrv,IEEEexample}

\begin{thebibliography}{10}
\providecommand{\url}[1]{#1}
\csname url@rmstyle\endcsname
\providecommand{\newblock}{\relax}
\providecommand{\bibinfo}[2]{#2}
\providecommand\BIBentrySTDinterwordspacing{\spaceskip=0pt\relax}
\providecommand\BIBentryALTinterwordstretchfactor{4}
\providecommand\BIBentryALTinterwordspacing{\spaceskip=\fontdimen2\font plus
\BIBentryALTinterwordstretchfactor\fontdimen3\font minus
  \fontdimen4\font\relax}
\providecommand\BIBforeignlanguage[2]{{%
\expandafter\ifx\csname l@#1\endcsname\relax
\typeout{** WARNING: IEEEtran.bst: No hyphenation pattern has been}%
\typeout{** loaded for the language `#1'. Using the pattern for}%
\typeout{** the default language instead.}%
\else
\language=\csname l@#1\endcsname
\fi
#2}}

\bibitem{Ziolkowski03}
R.~Ziolkowski, ``Application of double negative materials to increase the power
  radiated by electrically small antennas,'' \emph{IEEE Trans. Antennas
  Propag.}, vol.~51, no.~10, pp. 2626--2640, 2003.

\bibitem{Mahmoud04}
S.~F. Mahmoud, ``A new miniaturized annular ring patch resonator partially
  loaded by a metamaterial ring with negative permeability and permittivity,''
  \emph{IEEE Ant. Wireless Propag. Lett.}, vol.~3, pp. 19--22, 2004.

\bibitem{Tretyakov05}
S.~A. Tretyakov and M.~Ermutlu, ``Modeling of patch antennas partially loaded
  with dispersive backward-wave materials,'' \emph{IEEE Ant. Wireless Propag.
  Lett.}, vol.~4, pp. 266--269, 2005.

\bibitem{Qureshi05}
F.~Qureshi, M.~A. Antoniades, and G.~V. Eleftheriades, ``A compact and
  low-profile metamaterial ring antenna with vertical polarization,''
  \emph{IEEE Ant. Wireless Propag. Lett.}, vol.~4, pp. 333--336, 2005.

\bibitem{Itoh06}
C.-J. Lee, K.~M. K.~H. Leong, and T.~Itoh, ``Composite right/left-handed
  transmission line based compact resonant antennas for {RF} module
  integration,'' \emph{IEEE Trans. Antennas Propag.}, vol.~54, no.~8, pp.
  2283--2291, 2006.

\bibitem{Stuart06}
H.~R. Stuart and A.~Pidwerpedtsky, ``Electrically small antenna elements using
  negative permittivity resonators,'' \emph{IEEE Trans. Antennas Propag.},
  vol.~54, no.~6, pp. 1644--1653, 2006.

\bibitem{Ziolkowski06}
R.~Ziolkowski and A.~Erentok, ``Metamaterial-based efficient electrically small
  antennas,'' \emph{IEEE Trans. Antennas Propag.}, vol.~54, no.~7, pp.
  2113--2130, 2006.

\bibitem{Alu07}
A.~Alu, F.~Bilotti, N.~Engheta, and L.~Vegni, ``Subwavelength, compact,
  resonant patch antennas loaded with metamaterials,'' \emph{IEEE Trans.
  Antennas Propag.}, vol.~55, no.~1, pp. 13--25, 2007.

\bibitem{Ziol07}
R.~W. Ziolkowski and A.~Erentok, ``At and below the {C}hu limit: passive and
  active broad bandwidth metamaterial-based electrically small antennas,''
  \emph{IET Proc.~Microw.~Antennas Propag.}, vol.~1, no.~1, pp. 116--128, 2007.

\bibitem{Mosallaei08}
S.~Chadarghadr, A.~Ahmadi, and H.~Mosallaei, ``Negative permeability-based
  electrically small antennas,'' \emph{IEEE Ant. Wireless Propag. Lett.},
  vol.~7, pp. 13--17, 2008.

\bibitem{Erentok08}
A.~Erentok and R.~Ziolkowski, ``Metamaterial-inspired efficient electrically
  small antennas,'' \emph{IEEE Trans. Antennas Propag.}, vol.~56, no.~3, pp.
  691--707, 2008.

\bibitem{Hirvonen08}
M.~Hirvonen and J.-E. Sten, ``Power and {Q} of a horizontal dipole over a
  metamaterial coated conducting surface,'' \emph{IEEE Trans. Antennas
  Propag.}, vol.~56, no.~3, pp. 684--690, 2008.

\bibitem{Bilotti08}
F.~Bilotti, A.~Alu, and N.~Engheta, ``Design of miniaturized metamaterial patch
  antennas with $\mu$-negative loading,'' \emph{IEEE Trans. Antennas Propag.},
  vol.~56, no.~6, pp. 1640--1647, 2008.

\bibitem{Mittra1}
R.~Mittra, ``A critical look at metamaterials for antenna related
  applications,'' \emph{Journal of Communications Technology and Electronics},
  vol.~52, no.~9, pp. 1051--1058, 2007.

\bibitem{Ikonen_pres}
P.~Ikonen, ``Some reflections on the appearance of metamaterials in microwave
  engineering,'' presented at the {\it {Young Scientist Meeting on
  Metamaterials}}, 2008.

\bibitem{Vainikainen}
P.~Vainikainen, J.~Ollikainen, O.~Kivekas, and I.~Kelander, ``Resonator-based
  analysis of the combination of mobile handset antenna and chassis,''
  \emph{IEEE Trans. Antennas Propag.}, vol.~50, no.~10, pp. 1433--1444, 2002.

\bibitem{Ikonen}
P.~M.~T. Ikonen, S.~I. Maslovski, C.~R. Simovski, and S.~A. Tretyakov, ``On
  artificial magnetodielectric loading for improving the impedance bandwidth
  properties of microstrip antennas,'' \emph{IEEE Trans. Antennas Propag.},
  vol.~54, no.~6, pp. 1654--1662, 2006.

\bibitem{IkonenMOTL}
P.~Ikonen, S.~Maslovski, and S.~Tretyakov, ``{PIFA} loaded with artificial
  magnetic material: practical example for two utilization strategies,''
  \emph{Microw. Opt. Techn. Lett.}, vol.~46, no.~3, pp. 205--210, 2005.

\bibitem{Hansen}
R.~C. Hansen and M.~Burke, ``Antennas with magnetodielectrics,'' \emph{Microw.
  Opt. Techn. Lett.}, vol.~26, no.~2, pp. 75--78, 2000.

\bibitem{Brown}
J.~Brown, ``Artificial dielectrics,'' \emph{Progress in Dielectrics}, vol.~2,
  pp. 195--225, 1960.

\bibitem{Wheeler}
H.~A. Wheeler, ``The spherical coild as an inductor, shield, or antenna,''
  \emph{Proc. {IRE}}, vol.~46, pp. 1595--1602, 1958.

\bibitem{Stuart_conf}
H.~R. Stuart, ``An electromagnetic comparison of the tapered spherical helix
  and the negative permittivity sphere,'' in \emph{Proc. {IEEE} International
  Symposium on Antennas and Propag. '07}, Hawaii, USA, June 2007, pp.
  3472--3475.

\bibitem{Kildal06}
P.~S. Kildal, ``Comments on ``application of double negative materials to
  increase the power radiated by electrically small antennas'','' \emph{IEEE
  Trans. Antennas Propag.}, vol.~54, no.~2, p. 766, 2006 (Authors' reply: pp.
  766-767).

\bibitem{Mosallaei07}
H.~Mosallei and K.~Sarabandi, ``Design and modeling of patch antenna printed on
  magneto-dielectric embedded-circuit metasubstrate,'' \emph{IEEE Trans.
  Antennas Propag.}, vol.~55, no.~1, pp. 45--52, 2007.

\bibitem{Buell06}
K.~Buell, H.~Mosallei, and K.~Sarabandi, ``A substrate for small patch antennas
  providing tunable miniaturization factors,'' \emph{IEEE Trans. Microw. Theory
  Tech.}, vol.~54, no.~1, pp. 135--145, 2006.

\bibitem{Sergei_book}
S.~A. Tretyakov, \emph{Analytical modeling in applied electromagnetics}.\hskip
  1em plus 0.5em minus 0.4em\relax Norwood, MA: Artech House, 2003.

\end{thebibliography}

\end{document}